\documentclass[10pt,a4paper,onecolumn]{article}
\usepackage{marginnote}
\usepackage{graphicx}
\usepackage{xcolor}
\usepackage{authblk,etoolbox}
\usepackage{titlesec}
\usepackage{calc}
\usepackage{tikz}
\usepackage{hyperref}
\hypersetup{colorlinks,breaklinks,
            urlcolor=[rgb]{0.0, 0.5, 1.0},
            linkcolor=[rgb]{0.0, 0.5, 1.0}}
\usepackage{caption}
\usepackage{tcolorbox}
\usepackage{amssymb,amsmath}
\usepackage{ifxetex,ifluatex}
\usepackage{seqsplit}
\usepackage{fixltx2e} 
\usepackage[
  backend=biber,
]{biblatex}
\bibliography{paper.bib}

\usepackage[top=3.5cm, bottom=3cm, right=1.5cm, left=1.0cm,
            headheight=2.2cm, reversemp, includemp, marginparwidth=4.5cm]{geometry}

\titleformat{\section}
  {\normalfont\sffamily\Large\bfseries}
  {}{0pt}{}
\titleformat{\subsection}
  {\normalfont\sffamily\large\bfseries}
  {}{0pt}{}
\titleformat{\subsubsection}
  {\normalfont\sffamily\bfseries}
  {}{0pt}{}
\titleformat*{\paragraph}
  {\sffamily\normalsize}

\usepackage{fancyhdr}
\makeatletter
\let\ps@plain\ps@fancy
\fancyheadoffset[L]{4.5cm}
\fancyfootoffset[L]{4.5cm}

\definecolor{linky}{rgb}{0.0, 0.5, 1.0}

\newtcolorbox{repobox}
   {colback=red, colframe=red!75!black,
     boxrule=0.5pt, arc=2pt, left=6pt, right=6pt, top=3pt, bottom=3pt}

\newcommand{\ExternalLink}{%
   \tikz[x=1.2ex, y=1.2ex, baseline=-0.05ex]{%
       \begin{scope}[x=1ex, y=1ex]
           \clip (-0.1,-0.1)
               --++ (-0, 1.2)
               --++ (0.6, 0)
               --++ (0, -0.6)
               --++ (0.6, 0)
               --++ (0, -1);
           \path[draw,
               line width = 0.5,
               rounded corners=0.5]
               (0,0) rectangle (1,1);
       \end{scope}
       \path[draw, line width = 0.5] (0.5, 0.5)
           -- (1, 1);
       \path[draw, line width = 0.5] (0.6, 1)
           -- (1, 1) -- (1, 0.6);
       }
   }

\patchcmd{\@maketitle}{center}{flushleft}{}{}
\patchcmd{\@maketitle}{center}{flushleft}{}{}
\patchcmd{\@maketitle}{\LARGE}{\LARGE\sffamily}{}{}
\def\maketitle{{%
  
  \AB@maketitle}}
\makeatletter
\renewcommand\AB@affilsepx{ \protect\Affilfont}
\renewcommand\AB@affilnote[1]{{\bfseries #1}\hspace{3pt}}
\makeatother

\renewcommand\Affilfont{\sffamily\small\mdseries}
\ifnum 0\ifxetex 1\fi\ifluatex 1\fi=0 
  \usepackage[T1]{fontenc}
  \usepackage[utf8]{inputenc}

\else 
  \ifxetex
    \usepackage{mathspec}
  \else
    \usepackage{fontspec}
  \fi
  \defaultfontfeatures{Ligatures=TeX,Scale=MatchLowercase}

\fi
\IfFileExists{upquote.sty}{\usepackage{upquote}}{}
\IfFileExists{microtype.sty}{%
\usepackage{microtype}
\UseMicrotypeSet[protrusion]{basicmath} 
}{}

\usepackage{hyperref}
\hypersetup{unicode=true,
            pdftitle={easyaccess: Enhanced SQL command line interpreter for astronomical surveys},
            pdfborder={0 0 0},
            breaklinks=true}
\usepackage{graphicx,grffile}
\makeatletter
\def\maxwidth{\ifdim\Gin@nat@width>\linewidth\linewidth\else\Gin@nat@width\fi}
\def\maxheight{\ifdim\Gin@nat@height>\textheight\textheight\else\Gin@nat@height\fi}
\makeatother
\setkeys{Gin}{width=\maxwidth,height=\maxheight,keepaspectratio}
\IfFileExists{parskip.sty}{%
\usepackage{parskip}
}{
\setlength{\parindent}{0pt}
\setlength{\parskip}{6pt plus 2pt minus 1pt}
}
\ifx\paragraph\undefined\else
\let\oldparagraph\paragraph
\renewcommand{\paragraph}[1]{\oldparagraph{#1}\mbox{}}
\fi
\ifx\subparagraph\undefined\else
\let\oldsubparagraph\subparagraph
\renewcommand{\subparagraph}[1]{\oldsubparagraph{#1}\mbox{}}
\fi

\title{easyaccess: Enhanced SQL command line interpreter for astronomical
surveys}

        \author[1]{Matias Carrasco Kind}
          \author[2]{Alex Drlica-Wagner}
          \author[1]{Audrey Koziol}
          \author[1]{Don Petravick}
    
      \affil[1]{National Center for Supercomputing Applications, University of Illinois
at Urbana-Champaign. 1205 W Clark St, Urbana, IL USA 61801}
      \affil[2]{Fermi National Accelerator Laboratory, P. O. Box 500, Batavia,IL 60510,
USA}
  \date{\vspace{-5ex}}

\begin{document}
\maketitle

\marginpar{
  \sffamily\small

  {\bfseries DOI:} \href{https://doi.org/00.00000/joss.00000}{\color{linky}{00.00000/joss.00000}}

  \vspace{2mm}

  {\bfseries Software}
  \begin{itemize}
    \setlength\itemsep{0em}
    \item \href{http://joss.theoj.org/papers/}{\color{linky}{Review}} \ExternalLink
    \item \href{https://github.com/mgckind/easyaccess}{\color{linky}{Repository}} \ExternalLink
    \item \href{http://dx.doi.org/00.00000/zenodo.0000000}{\color{linky}{Archive}} \ExternalLink
  \end{itemize}

  \vspace{2mm}

  {\bfseries Submitted:} 00 January 0000\\
  {\bfseries Published:} 00 January 0000

  \vspace{2mm}
  {\bfseries License}\\
  Authors of papers retain copyright and release the work under a Creative Commons Attribution 4.0 International License (\href{http://creativecommons.org/lic
enses/by/4.0/}{\color{linky}{CC-BY}}).
}

\hypertarget{summary}{%
\section{Summary}\label{summary}}

\texttt{easyaccess} is an enhanced command line interpreter and Python
package created to facilitate access to astronomical catalogs stored in
SQL Databases. It provides a custom interface with custom commands and
was specifically designed to access data from the
\href{https://www.darkenergysurvey.org/}{Dark Energy Survey} Oracle
database, although it can easily be extended to another survey or SQL
database. The package was completely written in
\href{https://www.python.org/}{Python} and support customized addition
of commands and functionalities. Visit
\url{https://github.com/mgckind/easyaccess} to view installation
instructions, tutorials, and the Python source code for
\texttt{easyaccess}.

\hypertarget{the-dark-energy-survey}{%
\section{The Dark Energy Survey}\label{the-dark-energy-survey}}

The Dark Energy Survey (DES) (DES Collaboration 2005; DES Collaboration
2016) is an international, collaborative effort of over 500 scientists
from 26 institutions in seven countries. The primary goals of DES are
reveal the nature of the mysterious dark energy and dark matter by
mapping hundreds of millions of galaxies, detecting thousands of
supernovae, and finding patterns in the large-scale structure of the
Universe. Survey operations began on on August 31, 2013 and will
conclude in early 2019. For about 500 nights, DES has been taking
thousands of deep images of southern sky, which are transferred and
processed at the National Center for Supercomputing Applications
(\href{http://www.ncsa.illinois.edu/}{NCSA}). The images are processed
to produce catalogs of astronomical sources with hundreds of millions of
entries (billions in the case of individual detections), describing the
sources found within the images and other relevant metadata. A
significant subset of the DES data was recently
\href{https://des.ncsa.illinois.edu/releases/dr1}{made public} (DES
Collaboration 2018) and can be accessed through several mechanisms
including \texttt{easyaccess} and
\href{https://des.ncsa.illinois.edu/easyweb/}{web interfaces} that run
\texttt{easyaccess} as a backend. This public release includes
information for almost 400 million astrophysical sources and
complementary tables to allow scientific analysis.

\hypertarget{des-users}{%
\subsection{DES users}\label{des-users}}

The first release of \texttt{easyaccess} was on February 17th, 2015 and
since then, over 300 DES Collaborators have used it to access the DES
databases (Figure 1). We note that roughly 800 DES accounts exist, but
this includes all database users including those that created accounts
before the release of \texttt{easyaccess}. In August 2018 (version
1.4.4), we added support for the public DES data release, and since then
we have increased the number of public users.

\begin{figure}
\centering
\includegraphics{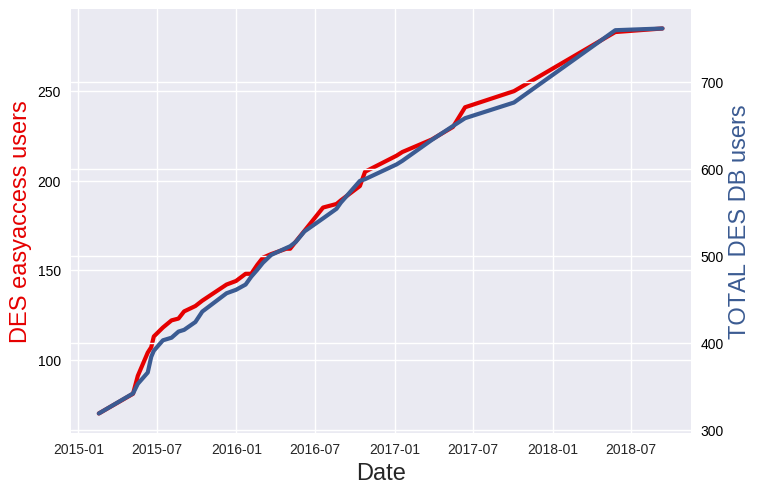}
\caption{Number of user since first version}
\end{figure}

\hypertarget{easyaccess}{%
\section{\texorpdfstring{\texttt{easyaccess}}{easyaccess}}\label{easyaccess}}

\texttt{easyaccess} is a command line interpreter that is heavily based
on \texttt{termcolor} (Lepa 2018) and the
\href{https://docs.python.org/3/library/cmd.html}{\texttt{cmd}} Python
core module. It interfaces with \texttt{cx\_Oracle} (Oracle Corp. 2018)
to communicate with Oracle, \texttt{requests} (Reitz 2012--2018) for
external URL requests, and other external open source libraries,
including NumPy (Oliphant 2006), \texttt{pandas} (McKinney 2010),
\texttt{fitsio} (Sheldon 2018) and \texttt{h5py} (Collette 2013) to
handle and transform array data. Figure 2 shows an example of the
welcome screen as seen by a DES user.

\begin{figure}
\centering
\includegraphics{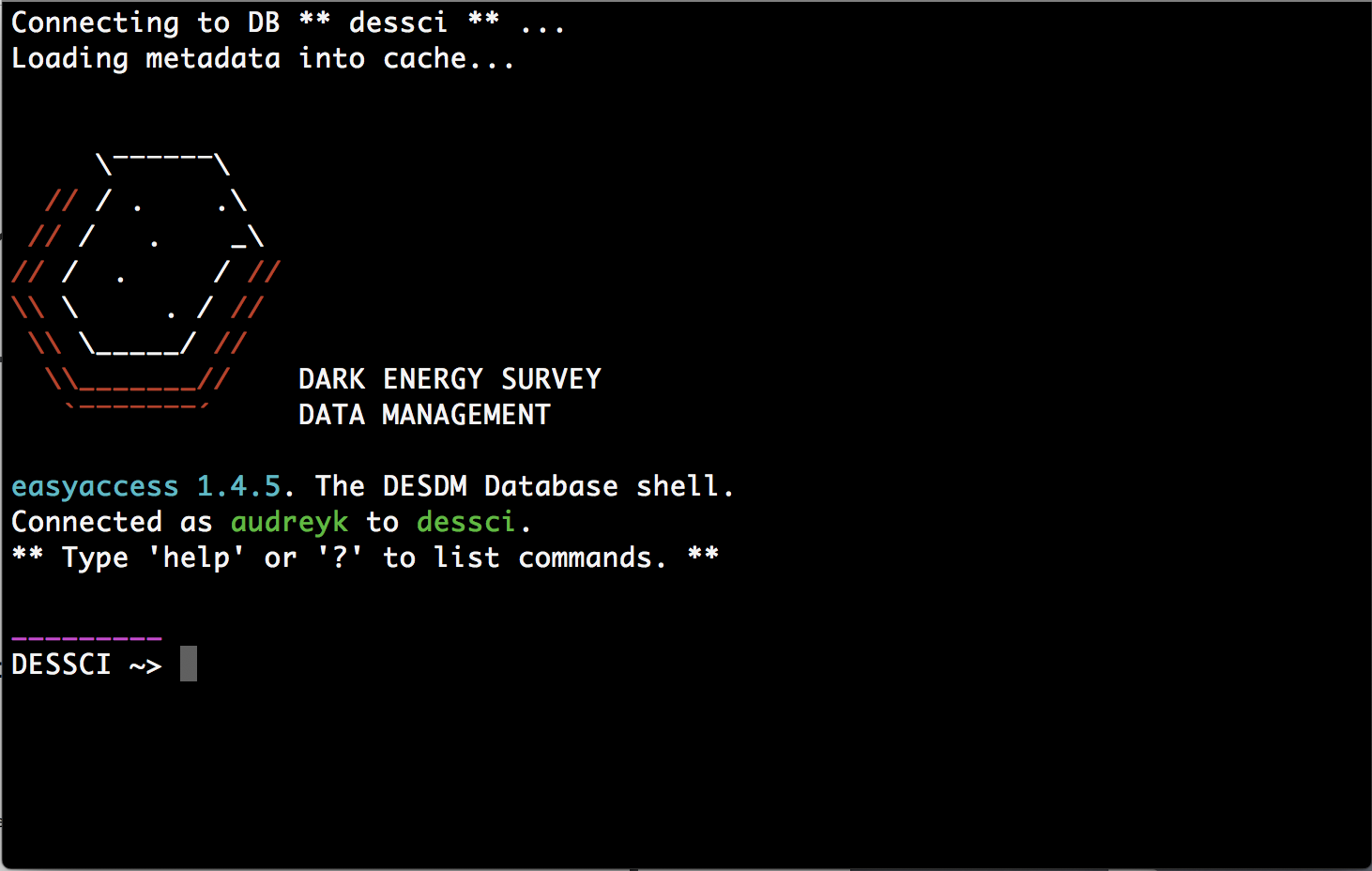}
\caption{Welcome screenshot}
\end{figure}

\hypertarget{features}{%
\subsection{Features}\label{features}}

\texttt{easyaccess} has a variety of features including a history of
past commands and smart tab auto-completion for commands, functions,
columns, users, tables, and paths. Tables can be written directly into
comma-separated-value (CSV) or white-space separated text files, FITS
(Wells, Greisen, and Harten 1981) files, and HDF5 (The HDF Group
1997--2018) files. It provides an iteration scheme to avoid memory
constraints when retrieving large tables. Tables can also be displayed
on the command line and most of the formatting is done using
\texttt{pandas}. Similarly, privileged users can easily upload tables to
the database from any of the file format described above in order to
share data with other users. The uploading mechanism is done chunk-wise,
allowing large tables to be loaded while keeping memory usage low.

In addition, there are a variety of customized functions to search and
describe the tables, search for users and user tables, check quota
usage, check the Oracle execution plan, and soon the ability to run
asynchronous jobs through a dedicated server. There are dozens of other
minor features that allow for a seamless experience while exploring and
discovering data within the hundreds of tables inside the DB.

One can also load SQL queries from a file into the database, or run SQL
queries inside the \texttt{easyaccess} python module in another IDE.
Most of the features are also exposed through a Python API and can be
run inside a Jupyter (Kluyver et al. 2016) notebook or similar tool
alongside scientific analyses.

Users can also use \texttt{easyaccess} to submit and request cutouts
around specific positions or objects which are generated from the
images. This allows better integration with other data services for a
richer scientific workflow.

\hypertarget{architecture}{%
\subsection{Architecture}\label{architecture}}

We have included a simplified UML diagram describing the architecture
and dependencies of \texttt{easyaccess} (Figure 3). Figure 3 shows only
the different methods for a given class and the name of the file hosting
a given class. The main class, \texttt{easy\_or()}, inherits all methods
from all different subclasses, making this model flexible and extendable
to other surveys or databases. These methods are then converted to
command line commands and functions that can be called inside
\texttt{easyaccess}. Given that there are some DES specific functions,
we have moved DES methods into a separate class \texttt{DesActions()}.

\begin{figure}
\centering
\includegraphics{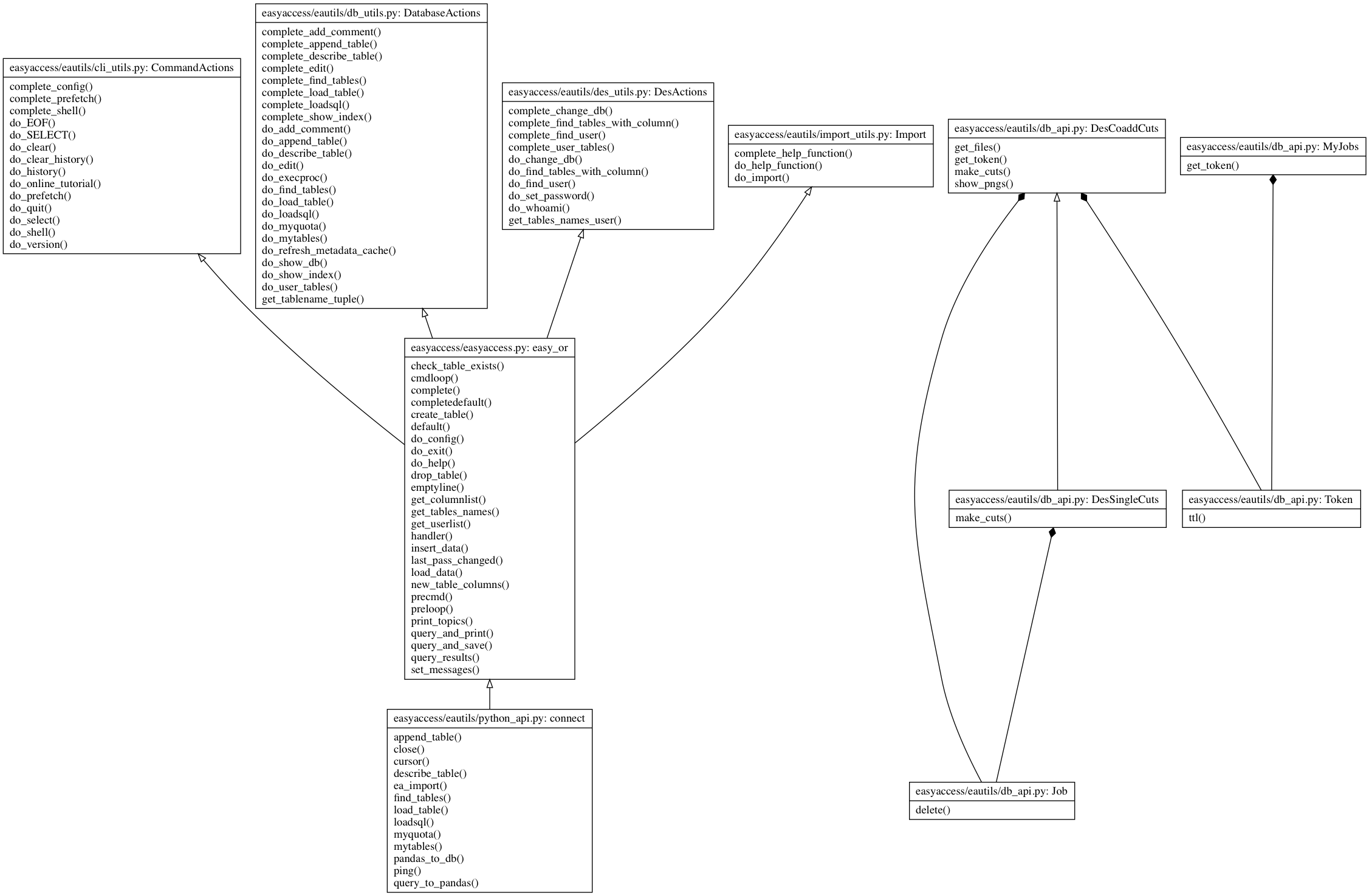}
\caption{\texttt{easyaccess} architecture diagram}
\end{figure}

\hypertarget{installation}{%
\subsection{Installation}\label{installation}}

To download easyaccess you can clone the source code from GitHub at
\url{https://github.com/mgckind/easyaccess} or follow any of the
standard installation channels described below.

\begin{itemize}
\item
  From \href{https://github.com/mgckind/easyaccess}{source}

  \texttt{python\ setup.py\ install}
\item
  \href{https://conda.io/docs/}{conda}

  \texttt{conda\ install\ easyaccess\ -c\ mgckind}
\item
  \href{https://hub.docker.com/r/mgckind/easyaccess/}{Docker}

  \texttt{docker\ pull\ mgckind/easyaccess}
\item
  \href{https://pypi.org/project/easyaccess/1.4.4/}{pip}

  \texttt{pip\ install\ easyaccess}
\end{itemize}

\hypertarget{acknowledgments}{%
\section{Acknowledgments}\label{acknowledgments}}

The DES Data Management System is supported by the National Science
Foundation under Grant NSF AST 07-15036 and NSF AST 08-13543.

\hypertarget{references}{%
\section*{References}\label{references}}
\addcontentsline{toc}{section}{References}

\hypertarget{refs}{}
\leavevmode\hypertarget{ref-h5py}{}%
Collette, Andrew. 2013. \emph{Python and Hdf5}. O'Reilly.

\leavevmode\hypertarget{ref-DES2005}{}%
DES Collaboration. 2005. ``The Dark Energy Survey.'' \emph{ArXiv
Astrophysics E-Prints}, October.

\leavevmode\hypertarget{ref-DES2016}{}%
---------. 2016. ``The Dark Energy Survey: more than dark energy - an
overview.'' \emph{MNRAS} 460 (August): 1270--99.
doi:\href{https://doi.org/10.1093/mnras/stw641}{10.1093/mnras/stw641}.

\leavevmode\hypertarget{ref-DR1}{}%
---------. 2018. ``The Dark Energy Survey Data Release 1.'' \emph{ArXiv
E-Prints}, January.

\leavevmode\hypertarget{ref-jupyter}{}%
Kluyver, Thomas, Benjamin Ragan-Kelley, Fernando Pérez, Brian Granger,
Matthias Bussonnier, Jonathan Frederic, Kyle Kelley, et al. 2016.
``Jupyter Notebooks -- a Publishing Format for Reproducible
Computational Workflows.'' Edited by F. Loizides and B. Schmidt. IOS
Press.

\leavevmode\hypertarget{ref-termcolor}{}%
Lepa, Konstantin. 2018. ``Termcolorr.'' \emph{PyPi Repository}.
\url{https://pypi.org/project/termcolor/}; PyPi.

\leavevmode\hypertarget{ref-pandas}{}%
McKinney, Wes. 2010. ``Data Structures for Statistical Computing in
Python.'' In \emph{Proceedings of the 9th Python in Science Conference},
edited by Stéfan van der Walt and Jarrod Millman, 51--56.

\leavevmode\hypertarget{ref-NumPy}{}%
Oliphant, Travis E. 2006. \emph{A Guide to Numpy}. Trelgol Publishing.

\leavevmode\hypertarget{ref-cxoracle}{}%
Oracle Corp. 2018. ``CxOracle.'' \emph{GitHub Repository}.
\url{https://github.com/oracle/python-cx_Oracle}; GitHub.

\leavevmode\hypertarget{ref-requests}{}%
Reitz, Kenneth. 2012--2018. ``Requests: HTTP for Humans.''
\url{http://docs.python-requests.org/}.

\leavevmode\hypertarget{ref-fitsio}{}%
Sheldon, Erin. 2018. ``Fitsio.'' \emph{GitHub Repository}.
\url{https://github.com/esheldon/fitsio}; GitHub.

\leavevmode\hypertarget{ref-hdf5}{}%
The HDF Group. 1997--2018. ``Hierarchical Data Format, version 5.''

\leavevmode\hypertarget{ref-FITS}{}%
Wells, D. C., E. W. Greisen, and R. H. Harten. 1981. ``FITS - a Flexible
Image Transport System'' 44 (June): 363.

\end{document}